\documentclass[epj]{webofc}
\usepackage[utf8]{inputenc}
\usepackage[varg]{txfonts}   
\usepackage{booktabs}
\usepackage{xcolor}
\definecolor{darkred}{rgb}{0.4,0.0,0.0}
\definecolor{darkgreen}{rgb}{0.0,0.4,0.0}
\definecolor{darkblue}{rgb}{0.0,0.0,0.4}
\usepackage[bookmarks,linktocpage,colorlinks,
    linkcolor = darkred,
    urlcolor  = darkblue,
    citecolor = darkgreen]{hyperref}
\usepackage{subfigure}
\wocname{EPJ Web of Conferences}
\woctitle{Lattice2017}
\usepackage{slashed}
\def\slashchar#1{\setbox0=\hbox{$#1$}           
   \dimen0=\wd0                                 
   \setbox1=\hbox{/} \dimen1=\wd1               
   \ifdim\dimen0>\dimen1                        
      \rlap{\hbox to \dimen0{\hfil/\hfil}}      
      #1                                        
   \else                                        
      \rlap{\hbox to \dimen1{\hfil$#1$\hfil}}   
      /                                         
   \fi}                                         %

\newcommand{\D}{\slashchar{D}}

\begin{document}
%
\selectlanguage{english}
\title{%
Numerical experiments using deflation with the HISQ action
}
\author{%
\firstname{Christine} \lastname{Davies}\inst{1} \and
\firstname{Carleton} \lastname{DeTar}\inst{2} \and
\firstname{Craig}  \lastname{McNeile}\inst{3}\fnsep\thanks{Speaker, \email{craig.mcneile@plymouth.ac.uk}} \and
\firstname{Alejandro}  \lastname{Vaquero}\inst{2} 
}
\institute{%
SUPA, School of Physics and Astronomy, University of Glasgow, Glasgow, G12 8QQ, UK
\and
Department of Physics and Astronomy,
University of Utah, Salt Lake City, UT 84112, USA
\and
Centre for Mathematical Sciences, Plymouth University, UK
}
\abstract{%
We report on numerical experiments using deflation
to compute quark
propagators for the 
highly improved staggered quark (HISQ) action.
The method is
tested on HISQ gauge configurations, generated by the MILC collaboration,
with lattice spacings of 0.15 fm, with a range of volumes, and 
sea
quark masses down to the physical quark mass. 
}
\maketitle
\section{Introduction}\label{intro}

An important goal of lattice QCD flavour physics calculations
is to find deviations from the predictions of the standard model
of particle physics. To exploit configurations with physical pion masses
requires speeding up the calculation of quark propagators and improved
measurement techniques to reduce statistical errors.
For example, the poor signal-to-noise ratio for the rho correlator at large times complicates the 
analysis of the hadronic vacuum polarization contribution to the 
anomalous magnetic moment of
 the muon~\cite{Chakraborty:2016mwy}. 
The eigenvalues and eigenvectors of the fermion matrix can
be useful both for speeding up the calculation of quark 
propagators and creating correlators with reduced errors.
The RBC collaboration 
has implemented
techniques that require thousands of eigenmodes to be 
calculated~\cite{Shintani:2014vja,CLehner}.

There has been much reported progress in developing faster algorithms for the
computation of quark propagators for both Wilson-like fermions and
domain wall (and overlap) fermions. 
The time taken to compute quark propagators has been speeded up
by factors of up to $O(15)$~\cite{Babich:2010qb,Frommer:2013fsa,Frommer:2013kla,Boyle:2014rwa},
using algorithms such as multigrid or domain decomposition,
applied to lattice QCD.

Another way to speed up an inversion is to remove eigenvalues
and eigenvectors from the matrix. This is known as exact deflation.
The use of deflation has been used to speed up the inversion
of sparse matrices~\cite{frank2001construction,saad2000deflated}
in applied mathematics.
Wilcox~\cite{Morgan:2004zh,Darnell:2007dr,AbdelRehim:2008nu}
has reviewed the various deflation 
algorithms~\cite{Wilcox:2007ei} used in lattice-QCD calculations developed
before 2007.
The xQCD collaboration~\cite{Li:2010pw}
have reported speed ups of 20 to 80, using deflation in combination with other techniques
for the calculation of quark propagators with the
overlap formalism on configurations with domain wall sea
quarks.

L\"{u}scher~\cite{Luscher:2007se,Luscher:2007es}
noted that the determination of the eigenvalues
for deflation has a potential cost that grows as  $O(V^2)$ for a volume $V$.
In principle inversion algorithms using multigrid or
domain decomposition should have a performance which
is independent of volume. In practice the parameter dependence of the algorithms
has to be tested.

There are close connections between 
matrix-inversion algorithms and
the algorithms used to compute 
eigenvalues~\cite{trefethen1997numerical}. For example,
Stathopoulos and Orginos have developed an algorithm,
called EIG-CG,  which 
combines sparse matrix inversion with the determination
of the eigenvalues~\cite{Stathopoulos:2007zi}. The original
algorithm worked for hermitian systems. They tested it 
with  anisotropic unquenched Wilson fermions. For single sources,
they found a speedup of between 7 and 10 for the ensembles they tested.
See~\cite{Cundy:2015jz} for a similar algorithm.

There has been much
less reported progress in the inversion algorithms for
staggered fermions. 
The staggered fermion operator is antihermitian, and thus the
inversion algorithm generally used is conjugate gradient 
for the normal equations, 
exploiting
even-odd
pre-conditioning. 
The Wilson-like operators are nonnormal, and additional algorithms were developed
to work with this type of matrix.
The calculation of the
inverse operator for overlap and domain wall uses a nested procedure.
The performance of deflation techniques for the inversion of staggered
operators could thus be very different to those for domain wall, overlap or
Wilson theories. 
The main development in
speeding up the inversion algorithms for staggered fermions was
use of multimass inverters~\cite{Jegerlehner:1996pm}.
After 20 years~\cite{Kalkreuter:1994ax} work has restarted on using
multigrid algorithms for staggered fermions~\cite{Weinberg:2017zlv,EWeinberg}, but 
algorithms are not yet available for QCD.

The determination of a large number of eigenmodes can be extremely costly.
One way to 
amortize
this startup cost is to 
reuse 
the eigenvectors 
many times, as, for example, 
by inverting the quark-fermion operator for multiple
right hand sides~\cite{AbdelRehim:2008nu}.
They can be 
used to compute
the low-mode contribution to
correlators from  all-to-all 
propagators~\cite{Neff:2001zr,DeGrand:2004qw,Foley:2005ac,Bali:2009hu,Bali:2015qya,Abdel-Rehim:2016pjw} 
treating only the high modes
with stochastic corrections. 
Also difficult-to-estimate disconnected loops 
can be computed using eigenvectors and eigenvalues.
Note that there are also newer techniques for reducing the noise~\cite{Ce:2016idq}
in lattice QCD calculations, which are not based on eigenmodes.

\subsection{Eigenvalues of the improved staggered Dirac operator}\label{sec-2}

Staggered Dirac operators, including 
the HISQ~\cite{Follana:2006rc}
operator, obey
\begin{equation}
\{ \slashed{D}  , \epsilon \}  = 0 \,,
\end{equation}
where
\begin{equation}
\epsilon = (-1)^{\sum_{\mu=1}^{d} x_\mu} \, .
\end{equation}
The massless staggered Dirac operator is antihermitian; hence,
the eigenvalue spectrum is purely imaginary.

\begin{equation}
\mathop{\mathrm{sp}}(\D) =
\{\pm i \lambda_s, \lambda_s \in \mathbb{R} \} \; .
\end{equation}
If $f_s$ is an eigenvector with eigenvalue $i \lambda_s$,
then $\epsilon \, f_s$ is also an eigenvector with $- i
\lambda_s$. 
There are potential zero modes $\lambda_s = 0$, related to
the topology of the gauge fields in the continuum. The lattice
approximation moves the eigenvalues from 0.

The fermion matrix for the HISQ action used in the inverter is
\begin{equation}
M = 
\begin{pmatrix}
m              & D_{eo}  \\
-D_{eo}^\dagger & m
\end{pmatrix}
\label{eq:staggM}
\end{equation}
where $D_{eo}$ is the part of the fermion operator which connects 
the even and odd sublattices and $m$ is the quark mass.

The following combination
\begin{equation}
D_{eo}D_{eo}^\dagger
\end{equation}
was used in the 
eigensolver. Unlike matrix inverters, the eigensolvers do 
not require that the matrix 
be positive definite.
The odd part of the j-th eigenvector,
($\chi_o^{j}$) \,,
can
be reconstructed from the even part ($\chi_e^{j}$) of the j-th eigenvector,
\begin{equation}
\chi_o^{j} = \frac{i}{\lambda^{j}} D_{oe}  \chi_e^{j} \,.
\end{equation}

\section{Eigensolvers in lattice QCD}\label{sec-1}

A key issue in the use of deflation is the performance of 
the algorithm used to determine the eigenvalues and eigenvectors.
The Lanczos  algorithm has been used from the very early
days of lattice QCD to determine eigenvalues. A three-
term recursion relation is used to find a tridiagonal matrix.
Unfortunately, the rounding errors generate  ghost eigenvalues.
There are many improvements of the basic Lanczos 
algorithm involving various types of restarts and vector spaces, which are implemented
in various software libraries or implemented in lattice QCD codes.
For example the MILC code contains routines which compute
eigenvalues using an accelerated conjugate gradient algorithm ~\cite{Kalkreuter:1995mm},
as well as an implementation of the EIG-CG algorithm~\cite{Stathopoulos:2007zi}.

There are a number of external sparse eigensolvers libraries commonly 
used in lattice QCD calculations. For example: the MILC code can call the
PRIMME (preconditioned iterative multi-method eigensolver methods)
library~\cite{stathopoulos2010primme}, which uses the Jacobi-Davidson 
method. 
For this project we added an interface to the 
ARPACK library~\cite{lehoucq1998arpack}, which uses
the implicitly restarted Arnoldi method (IRAM).
A comparison~\cite{Kalkreuter:1995vg} of a variant of the Lanczos algorithm
with the accelerated conjugate gradient algorithm ~\cite{Kalkreuter:1995mm},
found that Lanczos was better.
Most of our work on this project has used either the PRIMME or ARPACK libraries
to determine the eigenvalues.

The number of iterations required in solving linear equations
using conjugate gradient is related to the condition number
of the matrix. This condition number is crucial to understanding the
performance of the conjugate gradient algorithm, and indeed the idea
of exact deflation is based on reducing the condition number by removing
the lowest eigenvectors.

The condition number of an algorithm also expresses how sensitive
the output is to changes in the input.
There are condition numbers for the determination
of 
each
eigenvalues and eigenvectors.
Here we report some 
results (reviewed in~\cite{bai2000templates}
for the condition numbers of eigenvectors and eigenvalues
for Hermitian matrices.

If $x_i$ and $\lambda_i$ are estimates of the i-th eigenvector and eigenvalue
of the matrix $M$, respectively. Then a residual can be computed
\begin{equation}
r_j = \mid\mid M x_j - \lambda_i x_j \mid\mid_2 
\end{equation}
with the convention $\mid\mid x_j \mid\mid_2 = 1  $.
We use the notation $\hat{\lambda_j}$ and
$\hat{x}_j$
for the true eigenvalues and eigenvectors.
Then the error in the determination of a given eigenvalue is bounded:
\begin{equation}
\mid \lambda_j - \hat{\lambda_j} \mid \leq \mid\mid r_j \mid\mid_2
\label{eq:eigenErro}
\end{equation}

There is another relationship  based on how close together the eigenvalues are, which
may produce a tighter bound on the computed eigenvalues.
\begin{equation}
\delta_j = \min_{k \neq j} \mid \lambda_j - \lambda_k \mid
\end{equation}
\begin{equation}
\mid \lambda_j - \hat{\lambda_j} \mid \leq \frac{ \mid\mid r_j \mid\mid_2^2 }{\delta}
\end{equation}
The computed eigenvectors of a Hermitian matrix
are much more sensitive to input conditions than eigenvalues.
\begin{equation}
\sin \theta    \leq \frac{ \mid\mid r_j \mid\mid_2 }{\delta}
\end{equation}
The angle  $\theta$ is defined by
\begin{equation}
\hat{x_j} = x_j \cos \theta + z \sin \theta
\end{equation}
where $z$ is a vector orthognal to $x_j$.
This is the basis of the heuristic statement that it is cheaper to compute the
eigenvalue spectrum if the eigenvalues are widely separated.

\begin{figure}
\begin{minipage}{.5\textwidth}
\centering
\includegraphics[scale=0.35]{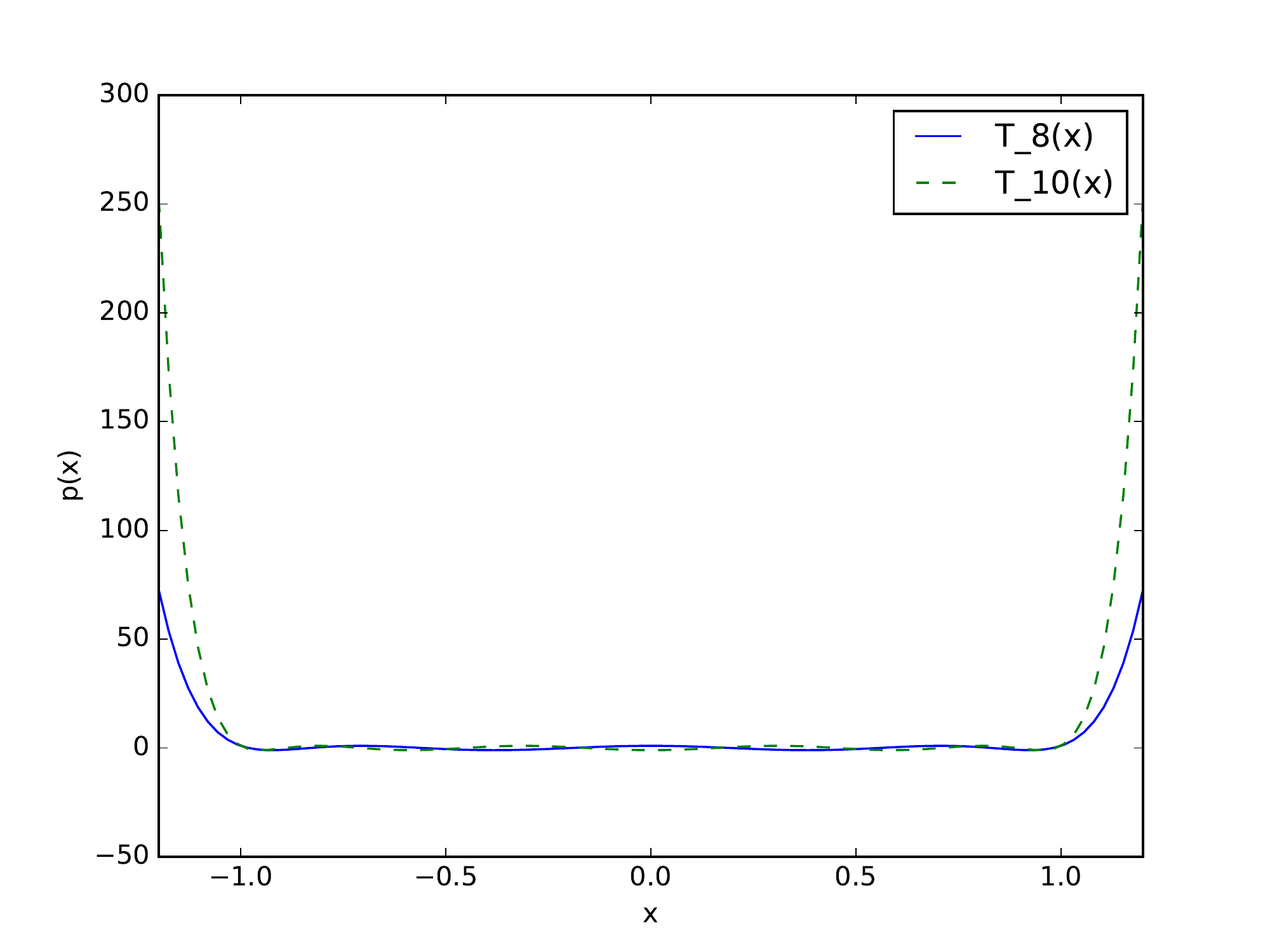}
\caption{Two Chebyshev polynomials\label{fig:ChebyPolyExample}}
\end{minipage}%
\begin{minipage}{.5\textwidth}
\centering
\includegraphics[scale=0.35]{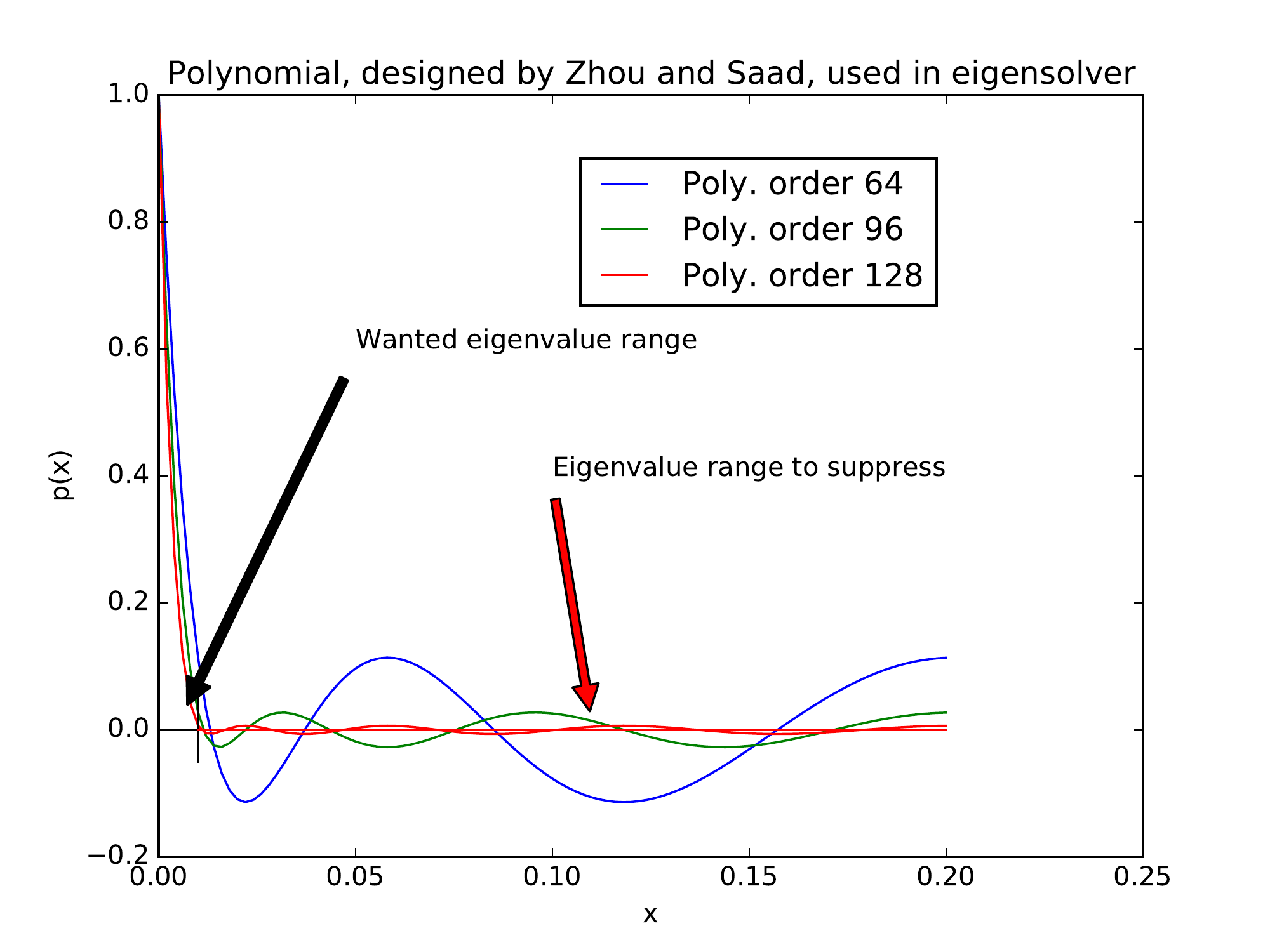}
\caption{Polynomials from Zhou and Saad~\cite{zhou2007chebyshev}\label{fig:ChebyPoly}}
\end{minipage}%
\end{figure}

To speed up the computation of the eigenvalues we are
experimenting with polynomial 
acceleration~\cite{Neff:2001zr,Morningstar:2011ka,Shintani:2014vja}
A polynomial of the matrix ($p(x)$) 
\begin{equation}
p(D_{eo}D_{eo}^\dagger)
\end{equation}
is used in the eigensolver. The idea is to suppress
the unwanted part of the eigenvalue spectrum and to spread
out the required part of the spectrum. Chebyshev polynomials
are a popular choice.
They
are of order 1 between -1 and 1, but 
grow rapidly
outside
this region. In the simplest scheme, the unwanted eigenvalues are mapped to lie between
-1 and 1. 
See the Chebyshev polynomials in figure~\ref{fig:ChebyPolyExample}.
Sorensen and Yang~\cite{sorensen1997accelerating}
have investigated using polynomial approximations to step 
functions in the eigensolver, but found that the use of Chebyshev
polynomials gave superior results to other polynomials they tested.

We have tested a polynomial proposed and 
investigated by Zhou and Saad~\cite{zhou2007chebyshev}.
Consider a matrix $M$ with eigenvalues~\cite{zhou2007chebyshev} 
in the range $[a_0,b]$.
The polynomial is designed to suppress the eigenvalues in the
range: $[a,b]$, where $a > a_0$. Define $e = \frac{(b-a)}{2}$ and
$c = \frac{(b+a)}{2}$.

The iterative scheme for the polynomials is below
\begin{eqnarray}
x_{j+1} & = & \frac{2}{e}(M - c I) x_j - x_{j-1}  \nonumber \\
x_1     & = & ( M - c I) x_0 
\end{eqnarray}
Zhou and Saad~\cite{zhou2007chebyshev} introduced a modification to 
the above iterative scheme. They introduce a scaling factor
$\rho_j= C_j(\frac{2}{e}(a_0 - cI))$, so that the eigenvalues in the range $[a_0,a ]$ are mapped 
close to 1, and the eigenvalues in the range $[a,b]$ are mapped close to 0.
In figure~\ref{fig:ChebyPoly}, we plot  polynomials of different orders suggested by
Zhou and Saad~\cite{zhou2007chebyshev} for target eigenvalues in the range
0.0 to 0,01.

\section{Results from numerical experiments}  \label{se:eigenspec}

The goal is to solve the linear equations 
\begin{equation}
M \underline{x} = \underline{b}
\end{equation}
for $\underline{x}$,
      given $\underline{b}$ and the fermion matrix $M$.
The conjugate gradient algorithm
      reduces the norm of the
residual ($r_i = M \underline{x_i} - \underline{b} $) at every iteration,
and hence finds $\underline{x}$.

We have investigated the performance of the eigensolver and
the deflated inverter using
($n_f$ = 2+1+1) HISQ ensembles 
generated by 
the MILC collaboration~\cite{Bazavov:2010ru,Bazavov:2012xda}.
Three ensembles, with lattice spacing of $a \sim 0.15 $fm, were investigated.
The volumes and pion masses of the three ensembles are:
$16^3 \times 48 $ and $m_\pi \sim$ 310 MeV; 
$24^3 \times 48 $ and $m_\pi \sim$ 220 MeV; 
and 
$32^3 \times 48 $ and $m_\pi \sim$ 134 MeV.

\begin{figure}
\begin{minipage}{.5\textwidth}
\centering
\includegraphics[scale=0.35]{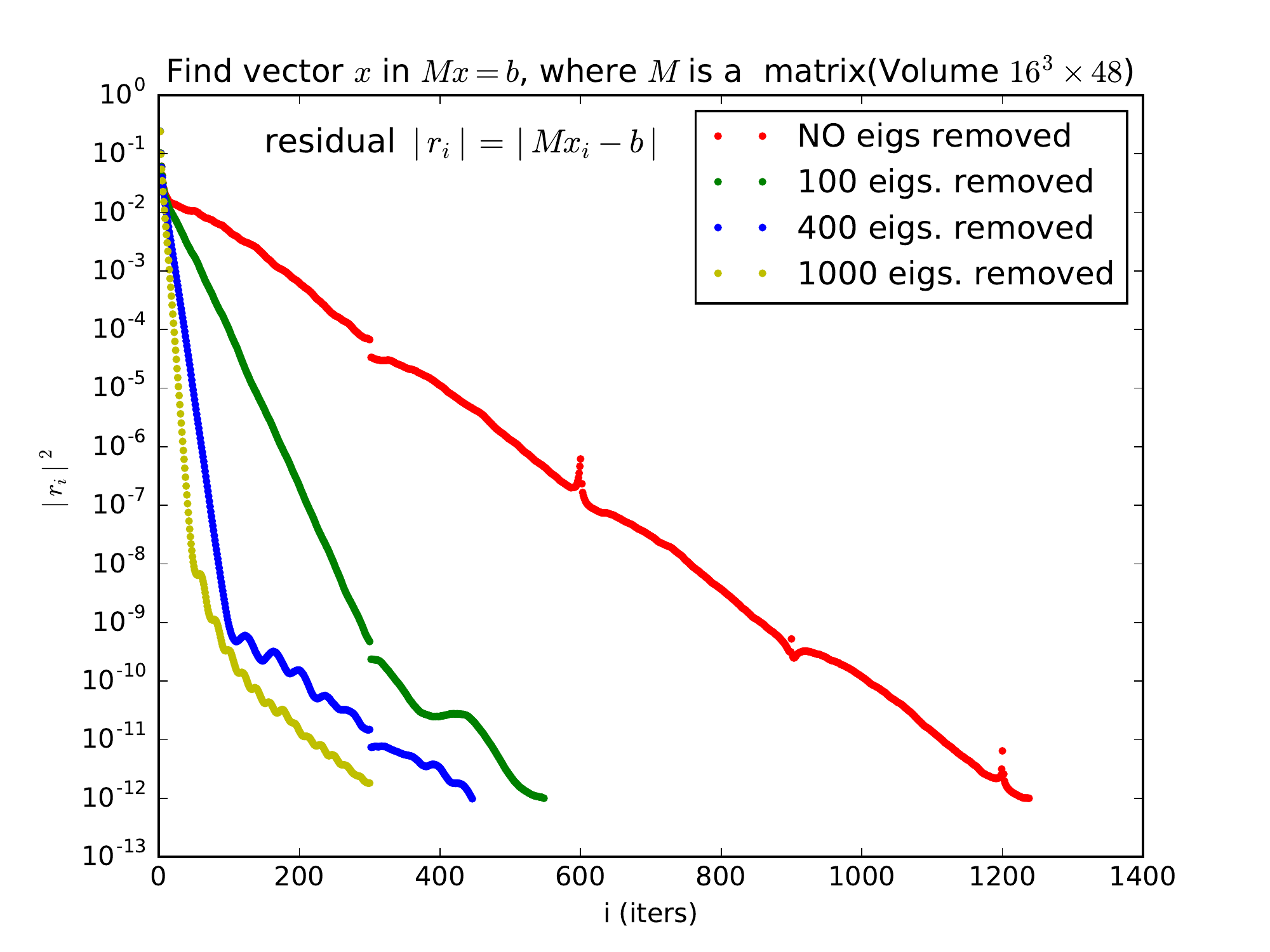}
\caption{Square of the residual as a function of iteration with a number of eigenvalues projected out. \label{fg:defRes}}
\end{minipage}
\begin{minipage}{.5\textwidth}
\centering
\includegraphics[scale=0.35]{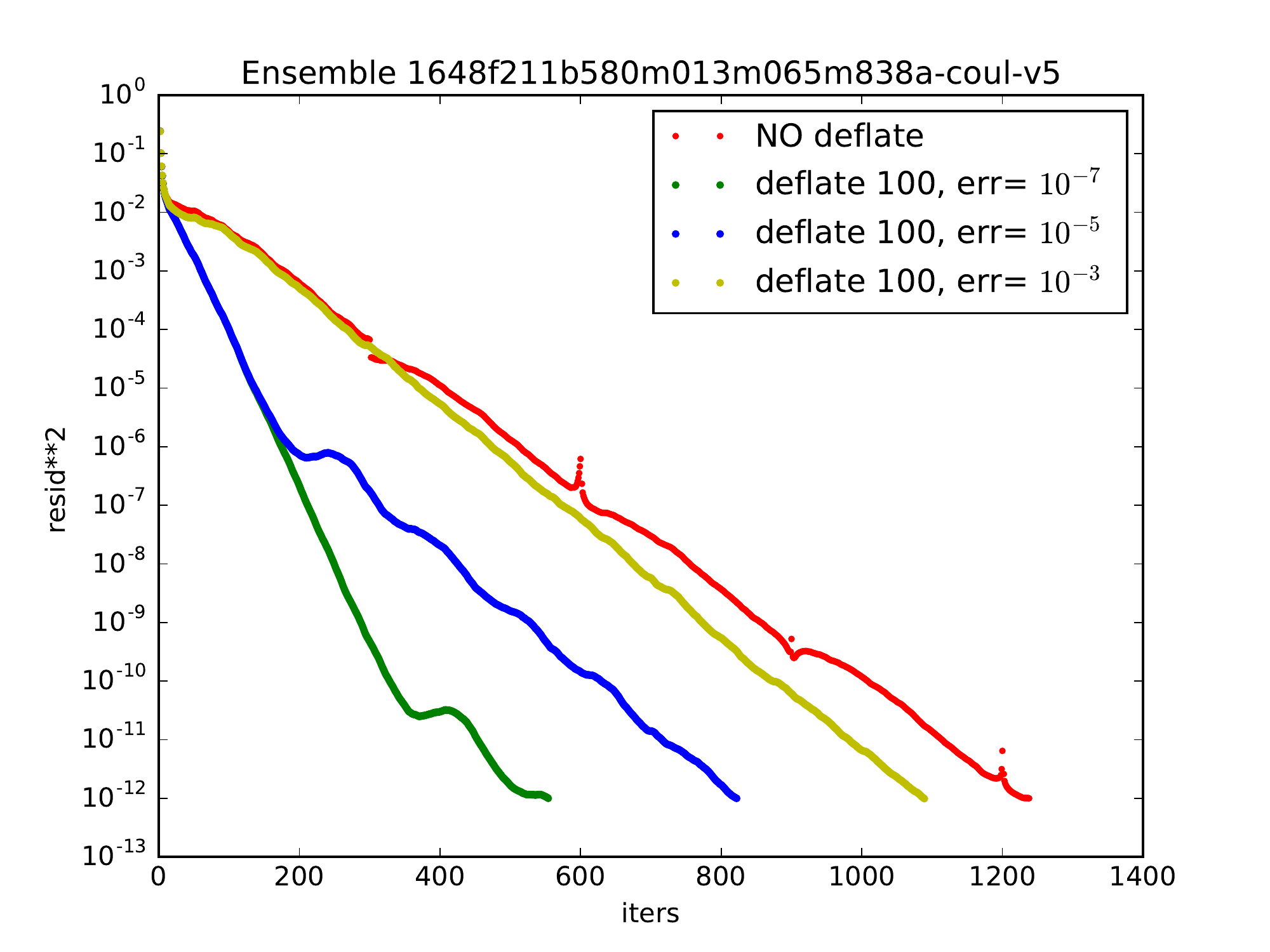}
\caption{Square of the residual as a function of iteration with a 100 eigenvalues projected out. 
The eigenvalues are determined with different accuracies.
\label{fg:defReseigen}}
\end{minipage}
\end{figure}

Figure~\ref{fg:defRes} shows the square of the residual of the CG inverter
as a function of the number of iterations 
as the number of deflated low eigenmodes is varied.
As expected, deflation reduces the number of
iterations required
to get to a 
given residual.
Progress stops when the residual reaches a value comparable the accuracy
      of the eigenmodes.
Figure~\ref{fg:defReseigen} shows the residual as a function of iteration when 100 eigenmodes are
deflated.
The eigenmodes are determined with 
varying
accuracy
using
the PRIMME eigensolver.

\begin{figure}
\centering
\includegraphics[scale=0.35]{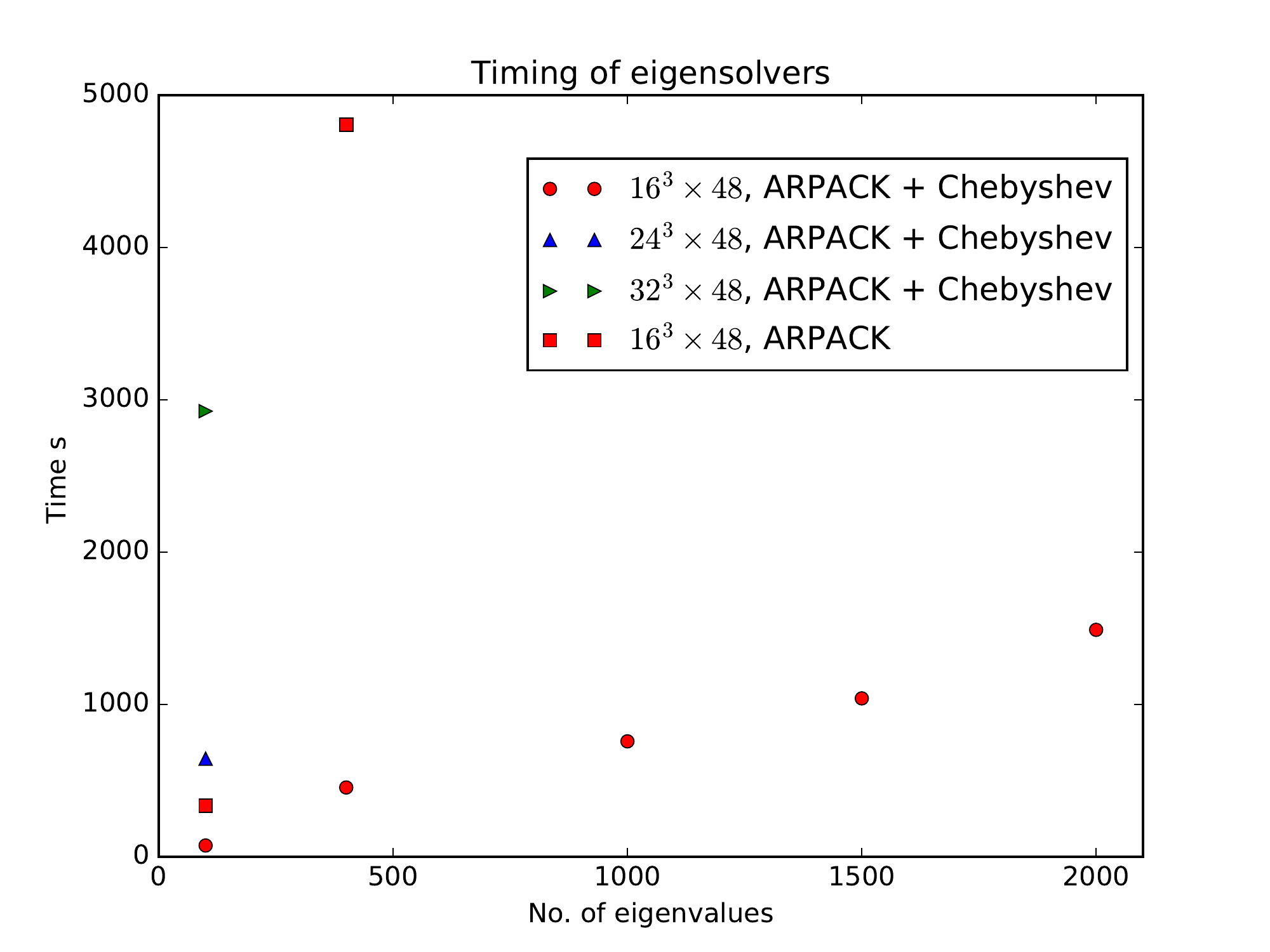}
\caption{Preliminary timings for eigensolvers on 64 cores \label{fg:time} for the 
ARPACK eigensolver.}
\end{figure}

An important issue is the performance of the eigensolver, because it
determines the setup cost for exact deflation.  The timings for the
ARPACK eigensolvers are plotted in figure~\ref{fg:time}.  We are still
tuning the Jacobi-Davidson algorithm in the PRIMME library. For example,
the paper~\cite{sorensen1997accelerating} uses polynomial acceleration
with a Davidson algorithm for a nonlattice QCD application.
We did not optimize the polynomials used in the acceleration
procedure. We stopped tuning the parameters 
of the
algorithm once we obtained 
residuals
around $10^{-14}$ with polynomial acceleration, so a further reduction of time
can probably be made.

Figure~\ref{fg:time} shows the preliminary time of determining different numbers of eigenmodes
with different algorithms. The tests were run on 64 nodes (Intel Sandy-Bridge cores).
The figure shows the dramatic speed up of the eigensolver when
polynomial acceleration is used. We are still tuning the various polynomials used and
the performance of the algorithm in the ARPACK library.
The figure also shows the increase in 
time to compute the eigenvalues
required as the volume is increased from $16^3 \times 48$ to $32^3 \times 48$.

\begin{figure}
\centering
\includegraphics[scale=0.35]{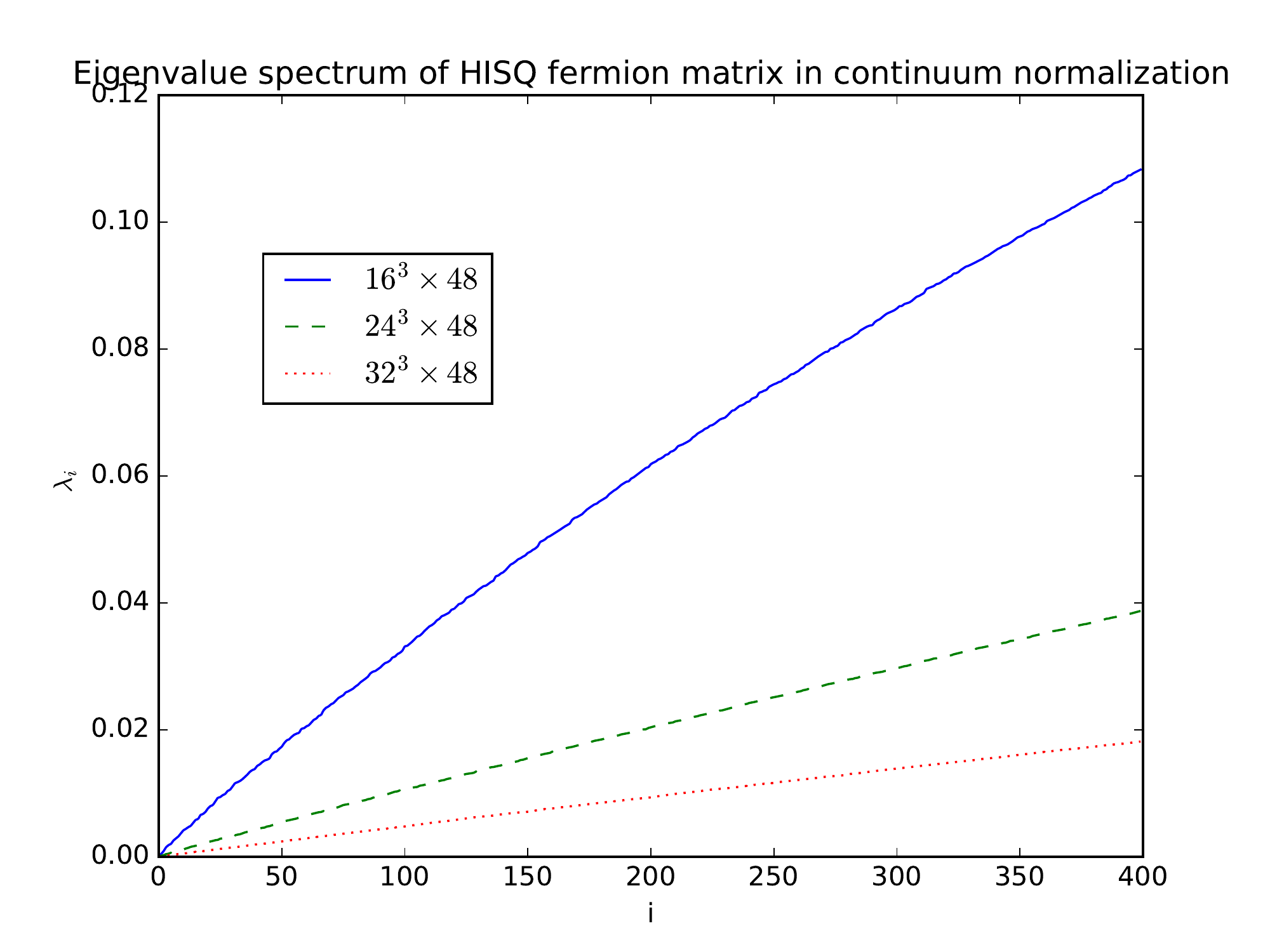}
\caption{Eigenvalue spectrum on the three ensembles~\label{fig:eigenspecVOL}}
\end{figure}

The performance of the eigensolver depends on the separation between the
eigenvalues, so it is useful to study the first O(1000) eigenvalues.
The majority (apart from~\cite{Catillo:2017qbz}.) of the studies of the eigenvalues of the
lattice Dirac operators has focused on the smallest eigenvalues,
because these are related to topology.

In figure~\ref{fig:eigenspecVOL} we plot the the eigenvalues 
of $D_{eo}$ (computed from the square root of the eigenvalues of
$D_{eo}D_{eo}^\dagger$)
on the
three volumes.
The magnitude
of the eigenvalues is approximately linear against the mode number.
A simple scaling of the bulk eigenvalues with $\frac{1}{V}$ for the space-time volume
$V$ doesn't map the eigenvalues onto a universal curve, so more study
is required of the volume dependence.
A log scale on the y-axis would 
help to reveal
the potential zero modes.
Follana et al.~\cite{Follana:2005km} argued that the near-zero topological eigenmodes
scale differently with volume than the bulk eigenmodes.

\section{Conclusions}  \label{se:conc}

There are many places in lattice QCD calculations where the
computation of thousands of eigenmodes are required, either in speeding
up the calculation of propagators, or in the design of better
measurement techniques. 
Our experience has been that a lot of tuning is required to get even
reasonable performance from an eigensolver.
It would be useful to have a better understanding~\cite{trefethen1997numerical} of how
to construct improved polynomials to use with an eigensolver.


We thank Carsten Urbach, 
Christoph Lehner, 
and Abdou Abdel-Rehim for discussions.
    This work used the Darwin Data Analytic system at the University
    of Cambridge, operated by the University of Cambridge High
    Performance Computing Service on behalf of the STFC DiRAC HPC
    Facility (www.dirac.ac.uk). This equipment was funded by a BIS
    National E-infrastructure capital grant (ST/K001590/1), STFC
    capital grants ST/H008861/1 and ST/H00887X/1, and DiRAC Operations
    grant ST/K00333X/1. DiRAC is part of the National
    E-Infrastructure.


\end{document}